# Structural transformation in supercooled water controls the crystallization rate of ice.

Emily B. Moore and Valeria Molinero*

*Department of Chemistry, University of Utah, Salt Lake City, UT 84112-0580, USA.*

**Contact Information for the Corresponding Author:**

VALERIA MOLINERO

Department of Chemistry, University of Utah

315 South 1400 East, Rm 2020

Salt Lake City, UT 84112-0850

Phone: (801) 585-9618; fax (801)-581-4353

Email: Valeria.Molinero@utah.edu

One of water's unsolved puzzles is the question of what determines the lowest temperature to which it can be cooled before freezing to ice. The supercooled liquid has been probed experimentally to near the homogeneous nucleation temperature $T_H \approx 232$ K, yet the mechanism of ice crystallization—including the size and structure of critical nuclei—has not yet been resolved. The heat capacity and compressibility of liquid water anomalously increase upon moving into the supercooled region according to a power law that would diverge at $T_s \approx 225$ K,[1,2] so there may be a link between water's thermodynamic anomalies and the crystallization rate of ice. But probing this link is challenging because fast crystallization prevents experimental studies of the liquid below $T_H$. And while atomistic studies have captured water crystallization[3], the computational costs involved have so far prevented an assessment of the rates and mechanism involved. Here we report coarse-grained molecular simulations with the mW water model[4] in the supercooled regime around $T_H$, which reveal that a sharp increase in the fraction of four-coordinated molecules in supercooled liquid water explains its anomalous thermodynamics and also controls the rate and mechanism of ice formation. The simulations reveal that the crystallization rate of water reaches a maximum around 225 K, below which ice nuclei form faster than liquid water can equilibrate. This implies a lower limit of metastability of liquid water just below $T_H$ and well above its glass transition temperature $T_g \approx 136$ K. By providing a relationship between the structural transformation in liquid water, its anomalous thermodynamics and its crystallization rate, this work provides a microscopic foundation to the experimental finding that the thermodynamics of water determines the rates of homogeneous nucleation of ice.[5]

Figure 1a presents the temperature dependence of some thermodynamics properties of liquid water as it is cooled at the lowest rate that still produces low-density amorphous ice (LDA) in simulations with the mW water model, which represents a water molecule as a single particle with short-range anisotropic interactions that mimic hydrogen bonds.[4] We note that in our simulations most molecules in LDA are four-coordinated as in ice albeit without long-range order,[6,7] and that the structural transformation from liquid water to amorphous ice is sharp but continuous. The sharp yet continuous nature of the structural transition agrees with the conclusions from experiments of water confined in narrow silica nanopores that prevent ice crystallization[8] and from a thermodynamic analysis[9] of bulk water outside the so-called "no-man's land" that stretches between the glass transition temperature $T_g$ and the homogenous nucleation temperature $T_H$.

The enthalpy of liquid water (top panel) decreases steeply around the liquid transformation temperature $T_L$=202±2 K (defined by the maximum change in density) and approaches the value for ice (Supplementary Figure 1). The heat capacity $C_p$ (middle panel) reaches a maximum at $T_L$, which is also the locus of maximum change in tetrahedrality and fraction of four-coordinated molecules.[7] We note that $T_L$ in the simulations is ~25K above $T_s$ estimated for real water, and ~15K above $T_s$ predicted by a power-law fit of $C_p$ values obtained with the mW water.[4] Large patches of four-coordinated molecules —the signature of LDA, and in an earlier simulation[3] identified as the precursor of the ice nucleus— develop in supercooled water and grow on cooling following a power law that would peak at $T_L$.[7], and that agrees with recent SAXS experiments of supercooled water down to 250 K.[10]

The development of crystallinity on water upon cooling is illustrated in Figure 1b. We note that calorimetry and x-ray diffraction report 5% ice in LDA[11], which is the same fraction we find in our simulations (Supplementary Figure 2). We also find that ice in LDA appears as small crystallites surrounded by threads of water with a local structure intermediate between ice and the four-coordinated liquid and without long-range order. This "intermediate-ice" structure accounts for ~20% of all the water present in LDA (Supplementary Figure 2) and may be a realization of the "gossamer percolative network" of nanocrystallites predicted to form at temperatures for which the length scale for motions relevant to the structural relaxation of the liquid is larger than the critical nucleus size for crystallization.[12] The large amount of threads of intermediate-ice that appear on approaching and crossing $T_L$ illustrates a blurring of the boundary between clearly liquid structures and clearly crystalline structures in deeply supercooled water. Below we show that this blurring heralds the effective limit of metastability of liquid water.

Experimentally observed crystallization rates increase when cooling the liquid towards $T_H$ but increase on heating the glass around $T_g$, implying the existence of a temperature $T_x$ of maximum crystallization rate in water's "no-man's land". The top panel of figure 1a indicates that on decreasing the rate of cooling, crystallization –evidenced by a sharp decrease in enthalpy– first occurs at $T_L$ and should thus be fastest at that temperature. Figure 2a gives the ice crystallization time $\tau_x$, computed from more than a thousand independent simulations, in the form of a time-temperature-transformation (TTT) diagram. The data show that $\tau_x$ is minimum at $T_x \approx 200$ K, almost identical to $T_L = 202 \pm 2$ K. For comparison, Figure 2b shows the TTT curve obtained when using nucleation theory and experimental data for water (Supplementary Discussion A): the crystallization time $\tau_x$ is predicted to be minimum and the crystallization rate maximum at $T_x \approx 225$ K, close to $T_s$ and just a few degrees below $T_H$.

To disentangle the contributions of structural transformation and degree of supercooling in determining $T_x$, we investigated the freezing of water that is confined in a 3 nm diameter cylindrical nanopore and therefore exhibits a decrease in the melting temperature of ice from 273 to 220 K for both the mW model[13] and experiment.[14] The simulations indicate that $T_L \approx T_x \approx 200$K in the pore (Supplementary Figure 3). We conclude that the freezing temperature of water is controlled by the structural transformation of the liquid and not merely the degree of supercooling. This explains the experimentally observed closing of the gap between freezing and melting of confined water on decreasing the radii of the confining nanopores.[14]

We also determined the number of water molecules $N^*$ in the critical ice nuclei (including their sheath of intermediate-ice) through the mean first passage time (MFPT) method.[15] Our values of $N^* \sim 120$ at $T_L+6$K and $\sim 90$ at $T_L+3$K are in good agreement with $N^*=70$-$210$ deduced from freezing of water in micelles around $T_H$.[16] The critical nucleus size is determined by liquid-ice thermodynamics, well reproduced by the mW model (Supplementary Discussions B and C). Critical nuclei around $T_L$+3K have a broad distribution of shapes (Figure 2c), evincing a lowering of the ice-liquid surface tension as the structural gap between ice and liquid narrows upon approaching $T_L$. The fraction of four-coordinated water molecules in the liquid wetting the nuclei at $T_L$+3K is 50% larger than the average for the whole liquid (Supplementary Figure 4), corroborating the observation of an earlier water freezing simulation study[3] that four-coordinated water patches that form in the supercooled liquid stabilize crystal nuclei.

The minimum in crystallization times around $T_L$ signals a crossover in the mechanism of ice crystallization, from nucleation-dominated above $T_L$ to growth-dominated below $T_L$ (Figure 2d). The lack of a well-defined nucleation plateau in the MFPT plot at $T_L$ (Supplementary Figure 5) is evidence of concurrent nucleation and growth, and implies that the barrier for nucleation is comparable to the thermal energy $RT$ (Supplementary Discussion D).[15] The growth time of the crystallites is comparable to or shorter than the relaxation time of the liquid because the growth rate is proportional to the diffusivity of supercooled liquid water,[17] which decouples from the structural relaxation[18] (Supplementary Discussions A and D). Thus liquid water cannot be equilibrated in the simulations at $T<T_L$+3K: ice nucleates before the liquid has time to equilibrate. We conclude that the structural transformation in supercooled water around $T_L$ sets the effective lower limit of metastability (LLM) of supercoooled water.

In fact, a kinetic limit of stability for the liquid state was anticipated by Kauzmann, in his seminal 1948 article,[19] as the resolution of his now famous entropy paradox: *"(…) the barrier to crystal nucleus formation, which tends to be very large just below the melting point, may at low temperatures be reduced to approximately the same height as the free energy barriers which impede molecular reorientations in the liquid (…). Under these circumstances crystal nuclei will form and grow at about the same rate as the liquid changes its structure following a change in temperature or pressure."* An extension of this argument considered the decoupling of diffusion and viscosity in supercooled liquids and concluded that a LLM must be reached at $T_{LLM} > T_o$, where the latter indicates the temperature at which the excess entropy of the liquid would become lower than the entropy of the crystal.[17] The experimental power law increase in $C_p$ of liquid water on cooling supports the existence of a LLM of water between $T_{LLM} > T_o \approx T_s \approx 225$ K and $T_H \approx 232$ K (Supplementary Discussions A and D); concurring with the prediction of a kinetic spinodal in water at ~230 K based on the fluctuation theory of relaxation of metastable states.[20]

The above arguments for real water, using classical nucleation theory and experimental data, predict a LLM close to the "nose" of the TTT curve, between $T_s$ and $T_H$—as predicted by our simulations. We therefore conclude that between $T_{LLM} \approx T_L > T_s$ and $T_g$, liquid water is not metastable and can only be studied over times shorter than needed for its equilibration. The low nucleation barriers and considerable water diffusivity around $T_{LLM}$ make partial crystallization unavoidable even at the fastest attainable cooling rates. Therefore the glass transition of LDA at $T_g \approx 136$ K does not produce metastable liquid water but a less viscous liquid unable to relax before crystallizing. The lack of ergodicity below ~225 K may explain the feeble heat capacity signature at $T_g$ that puzzled scientists for decades.[21]

Theoretical scenarios involving a retracing spinodal of superheated water, a first order and a continuous liquid-liquid transition have been proposed to explain the thermodynamic anomalies of water and predict its fate in "no-man's land."[21-25] These scenarios assume that metastable liquid water exists below $T_H$. The results of this work suggest that the structural transformation that causes the anomalies of water is also responsible for the demise of the liquid state: a low-density liquid (LDL) phase of water cannot be equilibrated. Water has been proposed to first convert to LDL and then crystallize[26], but our results reported here and in ref. [27] suggest that crystallization occurs faster than LDL's equilibration. It has been argued that LDL can be equilibrated in simulations with the ST2 model;[23,28] recently reported free energy maps of ST2 and mW models, however, do not display a basin for LDL.[29] We also note that our calculations indicate that water crystallization in "no-man's land" is limited only by the growth rate of the crystallites, which decreases on cooling. So extrapolation of crystallization rates from the nucleation-dominated region above $T_H$ to temperatures below 225 K, i.e., to temperatures relevant for cloud formation and crucial for the formulation of climate models, would severely overestimate the rates of ice formation.

**Methods Summary**

Simulations were performed with LAMMPS[30] using the mW water model,[4] which reproduces the structure, anomalies and phase behavior of water at less than 1% of the computational cost of atomistic models (Supplementary Discussion B). Thermodynamic properties of the liquid were computed as indicated in Methods Online for simulation cells containing 32,768 molecules at $p = 1$ atm and linear decrease in temperature at 10 K/ns. Ice was identified with the CHILL algorithm.[13] Crystallization times indicate average times to crystallize 70% of ~150 independent constant pressure and temperature simulations with 4,096 molecules at each temperature. Nucleation times and the average critical size and radius of gyration of the nuclei were determined from mean first passage time analysis[15] of the crystallization trajectories. Identification of individual critical nuclei was performed through evaluation of their individual crystallization probability over 200 independent simulations for each nucleus, starting with the same configuration and randomized momenta.

**Supplementary Information** is linked to the online version of the paper at www.nature.com/nature

**Acknowledgments.** This work was supported by the Arnold and Mabel Beckman Foundation through a Young Investigator Award to VM. We thank Pablo Debenedetti for discussions and Diego P. Fernandez for criticism of the manuscript.

**Author contributions.** VM conceived and designed the study and wrote the paper. EBM and VM performed the simulations, analyzed the data and interpreted the results.

**Author information.** Reprints and permissions information is available at www.nature.com/reprints. The authors declare no competing financial interests. Correspondence and requests for materials should be addressed to V.M. (Valeria.Molinero@utah.edu).

**FIGURE CAPTIONS.** *(470 words)*

**Figure 1. Evolution of the thermodynamics and structure of water on cooling. a)** Enthalpy $H$, heat capacity $C_p$ and excess free energy $G^{ex}$ of liquid mW water on hyperquenching to LDA glass at 10 K/ns (blue solid lines). Dashed line indicates the liquid transformation temperature $T_L$ which sets a lower limit of metastability (LLM) of liquid water. Thus $H$, $C_p$ and $G^{ex}$ for $T < T_L$ are not equilibrium quantities and depend on the cooling rate. Cooling mW water at 1K/ns results in crystallzation at $T_L$ (red line) where the crystallization rate is maximum. On cooling towards $T_L$, $C_p$ follows a power law (green dashed line)[4] with exponent 1.5 as in experiments[2] and $T_s$ about 35 K lower. $G^{ex}$ in the simulations (blue line) is in excellent agreement with $G^{ex}$ in experiments (solid cyan line).[9] The dashed cyan line shows the experimental $G^{ex}$ extrapolated below $T_H$.[5] **b)** Ice (red) and intermediate-ice (green) in: ice formed by cooling water at 1 K/ns (upper snapshot), LDA formed by quenching water at 10 K/ns (middle snapshot), and in liquid water at $T_L$ (lower snapshot). Lines connect water molecules within 3.5 Å.

**Figure 2. Kinetics of ice crystallization and critical ice nuclei. a)** Time-temperature transformation (TTT) diagram of mW water. Blue circles indicate average times $\tau_x$ to crystallize 70% of water. The error bars show large dispersion of crystallization times above $T_L$ due to the stochastic nature of the nucleation process. Water crystallizes within the nose (blue-shaded area). The crystallization time is the shortest around 200 K, close to $T_L=202\pm2$ K. Nucleation and growth times become comparable already at $T_L$ (Supplementary Figure 5). Above $T_L$, crystallization is limited by the rare formation of critical ice nuclei and supercooled water can be studied in the metastable liquid state. Below $T_L$, crystallization occurs before relaxation of liquid water. The maximum crystallization rate predicted by the mW model is $J=(\tau_x V)^{-1} \approx 10^{27} cm^{-3} s^{-1}$, several orders of magnitude faster than measured for water down to $T_H$, because crystallization rates are proportional to water mobility (section A of SI), overestimated by the mW model.[4,27] **b)** TTT curve predicted using Classical Nucleation Theory and experimental data for water. The nose resulting from crossover between nucleation and growth occurs at 225 K, between the experimental $T_H$ and $T_s$. **c)** Critical ice nuclei at the lowest temperature for which liquid water can be equilibrated, $T_L+3=205$ K in the simulations, contain ~90 water molecules and a wide range of compactness. The spread in $Rg$ of the nuclei (Supplementary Figures 6 and 7) suggests that the liquid-ice surface tension is very low on approaching $T_L$. **d)** Number of molecules in the largest ice nucleus for representative simulations at 208 K for which crystallization dominated by stochastic nucleation and at 192 K where nucleation is fast and crystallization proceeds at the pace of growth.

## METHODS ONLINE

**Simulations.** Molecular dynamics simulations were performed with LAMMPS.[30] Equations of motion were integrated using Velocity Verlet with a time step of 10 fs. Bulk simulations were conducted in the $NpT$ ensemble, with $p = 1$ atm. Temperature and pressure were controlled with the Nose-Hoover thermostat and barostat, with time constants of 1 and 5 ps, respectively. The target temperature was decreased linearly in the cooling ramps simulations. Water was modeled with the mW potential.[4] Three different systems were used in this study: i) The thermodynamics of bulk water was determined through cooling ramps with a simulation cell containing 32,768 mW water molecules. ii) The isothermal crystallization simulations were performed with cells containing 4,096 mW water molecules, after checking that it produced consistent results with a simulation cell containing 13,768 molecules. iii) The confined water system consisted of the 3 nm cylindrical nanopore employed in the studies of ref. 31 and contained 2123 water molecules embedded in a 5840 molecule pore. The interactions between pore-wall and water are chosen to be as water with water, to minimize the effect of the pore-wall on the liquid.[13] The pore was headless to ensure that the crystallization occurs within the shaft of the pore. The pore was 90% filled to allow for expansion of the water as it is cooled and forms ice. The nanopore simulations were performed in the *NVT* ensemble, although it should be noted that the water inside is at zero pressure as the pore is not fully filled.

**Thermodynamic properties.** The enthalpy, $H = E + pV$, was directly computed along the simulation trajectories and saved every 0.2 ps or less and averaged over 100 ps running intervals. The heat capacity was obtained through numerical differentiation of $H$ with respect to temperature. The excess entropy of liquid water with respect to ice (shown in Supplementary Figure 1) was obtained through integration of the change in entropy from the value at the melting point, $\Delta S_m = \Delta H_m / T_m$,

$$S^{ex}(T) = \Delta S_m(T_m) - \int_{T_m}^{T} \frac{c_p^{ex}(T')}{T'} dT' \tag{1}$$

where the excess heat capacity of liquid with respect to ice is $c_p^{ex}(T') = c_p^{liquid}(T') - c_p^{ice}(T')$. $C_p^{ice}(T)$ was computed in ref. 32, as well as the excess free energy, $G^{ex} = H^{ex} - TS^{ex}$, that here we extend down to 150 K. We note that at temperatures lower than 205 K the liquid cannot be equilibrated and the thermodynamic properties depend on the rate of cooling, which determines the fraction of water that crystallizes to ice.

**Identification of the Ice Nuclei.** The CHILL algorithm[13] was used to distinguish between molecules with local order of liquid, ice I, and molecules with local ordering intermediate between that on ice and liquid, that here we call intermediate-ice and we have called interfacial ice in refs.[13,27] because it is also formed on the interface between well defined crystallites and the liquid phase. An ice nucleus consists of clusters of molecules with any ice-like local environment, including both ice I and intermediate-ice. Ice nuclei are defined by clustering of ice and intermediate-ice molecules using 3.5 Å cutoff to define connected neighbors.

**Crystallization simulations at constant temperature.** To produce a large set (more than 100) of independent trajectories at each temperature, starting configurations were selected at 500 ps intervals from a single simulation at 300 K. From the starting configuration, the temperature was instantaneously quenched to the temperature of interest, $T_{quench}$, from 192 to 208 K, and the time was set to zero. The time required to convert 70% of the water into ice is considered the crystallization time, $\tau_X$. Nearly a thousand simulations up to 350 ns in length each were collected. The crystallization time of mW water at 180 K was taken from ref. [27].

**Nucleation times and critical nuclei size.** We used the mean first passage time (MFPT) method as implemented in ref. [15] to determine the characteristic timescale of nucleation. The number of water molecules in the largest ice nucleus, $N$, and its radius of gyration, $R_g$, were chosen as order parameters for the advance of the crystallization. The radius of gyration, $R_g$ and nonsphericity, NS of the nuclei were determined as described in ref. [7]. With the size of the ice nuclei as the order parameter, the nucleation time $\tau_{nuc}$ and the critical nuclei size, $N^*$ can be determined. For a series of trajectories at a given temperature, the mean time of first appearance is recorded for the largest nucleus in each configuration. A plot of the mean first passage time, the time it takes for a given nucleus size, $N$, to grow rather than dissolve for the first time, versus nuclei size results in a sigmoidal curve that can be described by the following equation:[15]

$$\tau(N_N) = \frac{\tau_N}{2}\left\{1 + erf\left[(N_N - N_N^*)c\right]\right\} \qquad (2)$$

where $\tau(n)$ is the mean first passage time as a function of cluster size, n. $\tau_{nuc}$ is the nucleation time and $N^*$ is the critical cluster size. The plateau of the sigmoidal curve gives the nucleation time, and the inflection point corresponds to the critical nucleus size $N^*$.

**Crystallization probability of individual nuclei.** A series of simulations were run to compare the growth probabilities of selected nuclei of size predicted to be critical by the MFPT method based on radius of gyration. Configurations containing a potentially critically sized nucleus, with size $N^*$, were chosen and the $R_g$ and nonsphericity of the nucleus was recorded. A series of 200 simulations were run from a set of independent configurations at 205 K, each initiated with newly randomized velocities, resulting in 200 unique simulations from each starting configuration. The probability of growth from the initial nucleus was calculated as the fraction of trajectories that resulted in nuclei growth after 5 ns, larger than the average growth time at 205 K, 2±1 ns.

**Local Liquid Environment of the Nuclei.** The liquid solvation shell of the crystalline clusters was analyzed for the crystallization trajectories at 200 and 205 K. The shell was comprised as the molecules of the liquid within 3.5 Å of any molecule of the crystal nucleus (the latter includes the intermediate-ice). We computed the ratio of four-coordinated ($N_4$) to higher-coordinated molecules ($N_H$) around the ice nuclei and compare it with the ratio for the whole system (that is about 40 times larger than the nuclei).

**Additional References for Methods.**

31. De La Llave, E., Molinero, V., and Scherlis, D.A., Water filling of hydrophilic nanopores. *J Chem Phys* **133** (3), 034513 (2010)

**Figure 1**

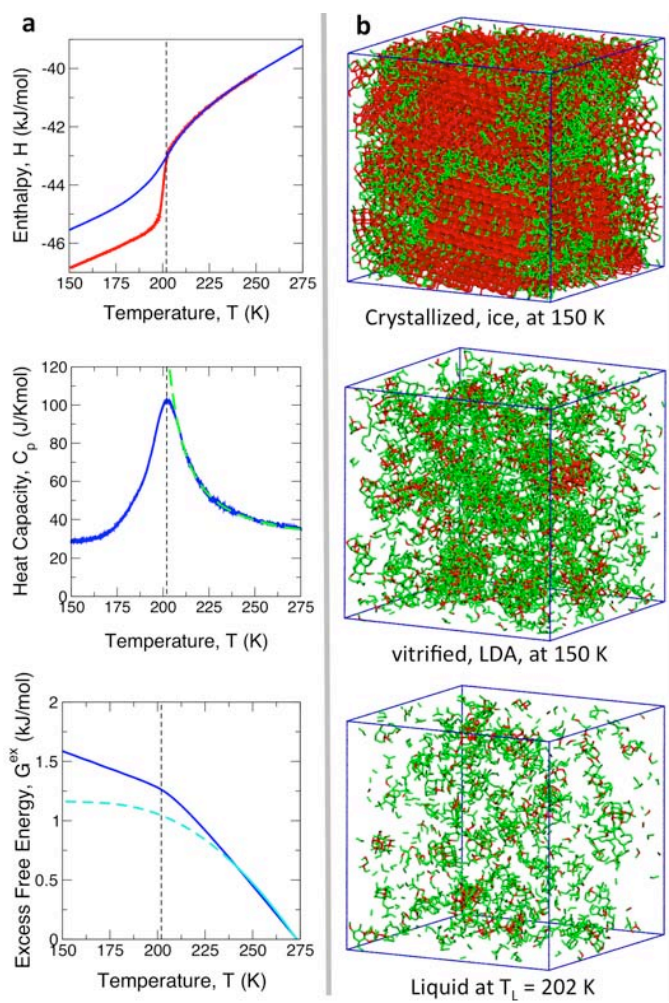

**Figure 2**

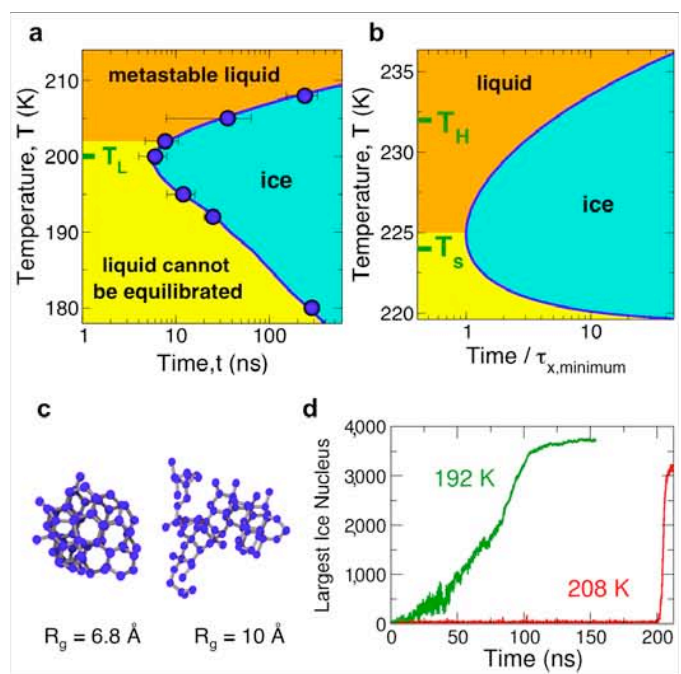

## SUPPLEMENTARY DISCUSSION
**A. On the effect of the enhanced diffusivity of the mW model on the crossover between nucleation-dominated and growth-dominated ice crystallization regimes.**

According to classical nucleation theory (CNT) the rate of crystal nucleation is the product of a pre exponential factor $C_N$ proportional to the rate of attachment $f$ of monomers (the water molecules) to the crystal nucleus and an exponential term that depends on the height of the nucleation barrier $\Delta G^*$ [(Kashchiev "*Nucleation: basic theory with applications*", Butterworth-Heinemann, New Delhi (2000)]:

$$I_N = C_N f \exp\left(-\frac{\Delta G^*}{kT}\right) \qquad (s.1)$$

$C_N$ has a weaker temperature dependence than the exponential term. The free energy barrier for nucleation is the reversible work of formation of the critical nucleus and depends only on equilibrium thermodynamic properties: the excess free energy of the liquid with respect to the crystal $G^{ex}$, the liquid-ice surface tension $\gamma$, and the number density of the crystal $\rho$:

$$\Delta G^* = c \frac{\gamma^3}{(\rho G^{ex})^2} \qquad (s.2)$$

where $c$ is a constant that depends on the geometry of the nucleus (e.g. $c = 16\pi/3$ for a spherical nucleus). The validity of the CNT equilibrium approximations, even under conditions close to a limit of stability, was confirmed in recent work [Maibaum, *Phys. Rev. Lett.*, **101**, 256102 (2008); Wedekind et al. *J. Chem. Phys.* **131**, 114506 (2009)].

The rate of attachment $f$ is proportional to the mobility of the molecules in the liquid, i.e. to the diffusion coefficient $D$ [Kashchiev "*Nucleation: basic theory with applications*", Butterworth-Heinemann, New Delhi (2000)]. The rate of growth is also proportional to $f$:

$$u_G = f\, C_G \left[1 - \exp(-G^{ex}/k_b T)\right], \qquad (s.3)$$

where $C_G$ is –to a good approximation- independent of temperature. It is important to note that experimental measurements of crystallization rates in atomic and molecular liquids indicate that *the crystallization rate is proportional to the translational diffusion* and not to the viscosity [Ngai et al., *J. Chem. Phys.*, **112**, 1887-1892 (2000); Swallen et al., *J. Phys. Chem. B*, 113, 4600 (2009)].

The crystallization time is given by $\tau_x = (3\phi/\pi I_N u_G^3)^{1/4}$, where $\phi$ is the volume fraction that crystallizes [Debenedetti, "*Metastable Liquids: concepts and principles*", Princeton University Press, Princeton (1996); Tanaka, *Phys. Rev. E.*, **68**, 011505 (2003)]. Writing explicitly the expression for the growth and nucleation rates in terms of thermodynamic quantities and the diffusional time $\tau_t$ (which is proportional to the rate of attachment $f$), the crystallization time can be expressed as [Tanaka, *Phys. Rev. E.*, **68**, 011505 (2003)]:

$$\tau_x = \tau_t\, g(\phi,T) = \tau_t\, C_\phi \Gamma(T) = \tau_t\, C_\phi \left( \frac{\exp(c\gamma^3/((\rho G^{ex})^2 RT))}{\left[1-\exp(-G^{ex}/RT)\right]^3} \right)^{1/4} \qquad (s.4)$$

where $g(\phi,T) = \tau_x/\tau_t$ is the ratio of the crystallization time for a fraction $\phi$ of the volume of the system and the characteristic translational time, and which can be represented as product of $C_\phi$ that depends on the fraction crystallized $\phi$ and is –to a good approximation- independent of temperature, and a term $\Gamma(T)$ that contains the temperature dependence and depends only on thermodynamic quantities. Equation s.4 indicates that the crystallization time is the product of a term that depends entirely on thermodynamic properties and a purely kinetic term that is inversely proportional to the characteristic diffusional time of water. In section B below we show that the mW model reproduces the thermodynamic properties of water that determine $\Gamma(T)$. The diffusion coefficient of mW, however, is larger than in experiment because the monatomic water model evolves on a smoother potential energy surface than fully atomistic water [Molinero and Moore *J. Phys. Chem. B*, *113*, 4008–4016 (2009); Moore and Molinero *J. Chem. Phys.* *132*, 244504 (2010)]. As a result, the crystallization rates at the nose of the TTT curve in the mW simulations are several orders of magnitude faster than would be in experiments or atomistic simulations. We take advantage of this high crystallization rate to study the crystallization by brute force molecular dynamics simulations.

The **temperature of maximum ice crystallization rate in real water** can be estimated from evaluation of $\Gamma(T)/D(T)$, which contains the temperature dependence of equation s.4 and thus is proportional to $\tau_x(T)$. Figure 2b shows $\tau_x(T)/\tau_{x,\text{minimum}}$ determined as $\Gamma(T)/D(T)$ divided by its minimum value. The expression for the diffusion coefficient $D(T)$ was taken from the work of Price et al. [*J. Phys. Chem. A*, *103*, 448-450 (1999)] which shows that a power law with singular temperature 219.2 K provides the best fit to the experimental diffusion coefficient of supercooled water. The thermodynamic term $\Gamma(T)$ was determined using the experimental $G^{ex}(T)$ [Koop et al, *Nature*, 406, 611-614 (2000)] and assuming that the surface tension has the same value at the melting point; the assumption of a constant surface tension as well as using the radius of the ice nucleus as a reaction coordinate would overestimate the barrier for nucleation (see section D below). Equation s.4 with the experimentally derived equations for $G^{ex}(T)$ and $D(T)$ predicts that the crystallization rate is maximum around 224.5 K, between the values of $T_H$ and $T_s$ for real water. See section D below for further discussions.

**B. On the accuracy of the mW model in predicting thermodynamic quantities of water relevant for ice crystallization.**

We assess now the accuracy of the mW model in predicting the thermodynamic quantities relevant for the nucleation and growth of ice, contained in the $\Gamma(T)$ term of equation s.4. The excess free energy $G^{ex}$ of supercooled liquid water with respect to ice is presented in Figure 1 for both the mW model (solid dark blue line) and water in experiments (cyan line). Actual measurements of the thermodynamics of liquid water exist only down to 235 K (solid cyan line); the lower temperature region of real water (dashed cyan line) was interpolated by Johari et al. from that liquid data and (non-equilibrium) data for LDA [*J. Phys. Chem.* 98, 4719-4725 (1994)] and later fitted to an analytical function of temperature by Koop et al [*Nature*, 406, 611-614 (2000)]. The agreement between the simulations and the experiments is excellent down to the lowest temperature for which there are actual measurements for water. We are not aware of reports of $G^{ex}$ as a function of supercooling for any other model of water.

We note that the mW model was parameterized to reproduce the experimental ratio of the excess enthalpy and entropy at the melting temperature of water: $T_m=\Delta H_m/\Delta S_m$. The actual values of $\Delta H_m$ and $\Delta S_m$ in the model are 12% lower than in the experiment [5.3 vs 6 kJ/mol and 19.3 vs 22 J/Kmol, respectively [Molinero and Moore, *J. Phys. Chem. B,* 113, 4008–4016 (2009)]. To place these results in perspective, we note that the level of agreement of mW with the experiments is significantly higher than for any of the commonly used atomistic water models (SPC, SPC/E, TIP3P, TIP4P, TIP5P), and as good or better than for TIP4P/2005 and TIP4P/ice, the atomistic potentials that best reproduce the phase diagram of water [Abascal and Vega, *J. Chem. Phys.*, 123, 234505 (2005)].

While the mW model does not have explicit hydrogen atoms, it reproduces the experimental pair and orientational distribution functions between oxygen atoms in liquid water [Molinero and Moore, *J. Phys. Chem. B.* 113, 4008 (2009)]. Thus the missing contribution to the entropy of mW water is the one arising from different configurations for the hydrogen atoms at fixed oxygen configurations. This should have two contributions: First, a configurational contribution due to proton disorder, that also exists in the crystal. The residual entropy of ice has been determined experimentally to be 3.43 J/Kmol [Giauque and Stout, *J. Am. Chem. Soc.*, 58, 1144 (1936)]; a comparable configurational contribution due to proton disorder may be expected for liquid water and LDA. Second, a contribution arising from the rotational freedom of the hydrogen atoms in liquid water. To gauge these contributions we note that the enthalpy of melting of ice (6.00 kJ/mol) is 12% of the enthalpy of sublimation of ice (51.06 kJ/mol). As the process of ice sublimation involves the breaking of two hydrogen bonds per mol of water, the low enthalpy of melting suggests that these hydrogen bonds are mostly intact in the liquid resulting in low orientational entropy of water in the liquid phase.

The excess entropy of liquid water with respect to ice (i.e. the melting entropy) is well reproduced by mW: the monatomic model underestimates the entropy of melting of ice $\Delta S_m$ by 12% [Molinero and Moore, *J. Phys. Chem. B.* 113, 4008 (2009)] while the best atomistic rigid models of water, TIP4P/2005 and TIP4P/ice underestimate $\Delta S_m$ by 13% and 11%, respectively. The other most commonly used models of water fare worse than the monatomic model: TIP4P underestimates $\Delta S_m$ by 14%, TIP4P/EW by 19% and TIP5P overestimates it by 21% [Abascal and Vega, *J. Chem. Phys.*, 123, 234505 (2005)]. Of all the atomistic models mentioned in this paragraph only mW and TIP4P/ice reproduce the melting temperature of ice (TIP4P, for example, underestimates $T_m$ by 41 K [Abascal and Vega, op. cit]). If the proton disorder contribution is similar for ice and the liquid or LDA, then the contribution of the rotational entropy to liquid water may be about 2.5 J/Kmol, the difference between the value predicted by mW and the experiment.

For completeness, we present the temperature dependence of the excess entropy $S^{ex}$ of the mW liquid with respect to ice (as determined by equation 1 of methods) in Supplementary Figure 1. Note that $S^{ex}(T)$ is the slope of $-G^{ex}(T)$ and the experimental and mW $G^{ex}(T)$ curve are almost identical down to 235 K, but the slope is larger for mW in the LDA region. The mW model predicts the excess entropy of LDA with respect to ice to be about 5.2 J/Kmol at 150 K. This excess entropy of the amorphous phase in the simulations is *higher* than that inferred from the analysis of the experimental vapor pressure of thin films of LDA at 150 K, $S^{ex}$ = 1.6±1 J/Kmol [Smith et al. *J. Phys. Chem. A.* 115, 5908 (2011)]. This indicates that at very low temperatures, below 240 K, the entropy of real LDA is even closer to the entropy of ice than predicted by the mW simulations. As the monatomic model does not have contributions from proton disorder or from rotational entropy, this result suggests that the lower excess entropy of LDA in experiments is due to more extensive crystallization than in the simulations. Widespread crystallization in the form of nanoscopic crystallites and threads of intermediate-ice may be undetected by conventional experimental methods. We caution, however, than thermodynamic quantities are ill defined at 150 K (and at any temperature below the limit of metastability of water) because liquid or amorphous water is not in thermodynamic equilibrium.

We have previously shown that simulations of liquid-ice equilibrium in cylindrical nanopores with the mW model reproduce the experimental melting temperatures and yield a Gibbs-Thomson constant for the liquid-ice equilibrium $K_{GT}$ = 54 K nm [Moore et al., *Phys. Chem. Chem. Phys.*, *12*, 4124-4134 (2010)] in excellent agreement with the theoretical values determined from water properties $K_{GT} = 2T_m\gamma_{sl}/(\rho\Delta H_m)$ = 52 K nm, and the 49.5 to 52 K nm measured in experiments. We note that $T_m$ is quantitatively reproduced by the mW model, the density of ice in mW is 6% higher than in experiments and $\Delta H_m$ is 12% lower than in experiments. Thus, the excellent agreement in the liquid-ice $K_{GT}$ suggests that the liquid-ice surface tension of the mW model is within 5-10% of the experimental value. Similarly, the surface tension between liquid and clathrate hydrates predicted by the mW model is in quantitative agreement with the values deduced from experiments [Jacobson and Molinero, *JACS*, *113*, 6458–6463, (2011)]. All these evidences indicate that the mW model properly predicts the thermodynamic properties relevant for the crystallization of water.

### C. Comparison of the critical ice nucleus size predicted by the simulations with the mW model with estimations from the literature.

The size of the critical nucleus in CNT results from a competition of the thermodynamic driving force ($G^{ex}$) and the free energy cost of producing an interface (proportional to the nucleus area and the liquid-ice surface tension γ). Assuming that the ice nucleus is spherical, the critical radius would be given by:

$$R_c = \frac{2\gamma}{\rho G^{ex}} \qquad (s.5)$$

In this work we determine that the $R_c$ at 205 K is about 1 nm and contains $N^* \approx 100$ water molecules. The non-compact nature of the critical ice nuclei at 205 K (Figure 2 and Supplementary Figure 7) indicates that the surface tension is lower than would be predicted by equation s.5. Based on the decrease of $G^{ex}$ with temperature (Figure 1), we estimate that $R_c$ would be ~50% larger at 230 K. The size predicted by the simulations is in good agreement with estimations or indirect determinations previously reported in the literature: Liu et al. interpreted the results of freezing of nanoscopic water droplet in emulsion with a CNT extended to confined systems and concluded that $N^* \approx$ 70 to 210 for homogeneous nucleation at 225 to 232 K [*Langmuir*, *23*, 7286-7292 (2007)]; using CNT and extrapolating thermodynamic properties for supercooled water to 230 K, Kashchiev predicted $N^* \approx$ 175 ["*Nucleation: basic theory with applications*", Butteworth-Heinemann, 2000]; the single crystallization simulation presented by Matsumoto and Ohmine is not sufficient to determine a critical nucleus size, nevertheless the fluctuations of crystallite size in the displayed trajectory at 230 K [*Nature*, *416*, 409-413 (2002)] suggest that $N^*$ may be around 150 in their simulation of TIP4P water.

### D. On the existence and location of a lower-limit of metastability in real water.

Kauzmann's solution of the entropy paradox considered only the magnitude of the relaxation and nucleation free energy barriers, showing that the two would become comparable before reaching the temperature $T_o$ at which the excess entropy of the liquid becomes zero. The heat capacity of supercooled liquid water increases following a power law with singular temperature $T_s \approx$ 225 K down to the lowest temperatures measured in experiments [Speedy and Angell, *J. Chem. Phys.*, *65*, 851-858 (1976); Tombari et al., *Chem. Phys. Lett., 300*, 749-751 (1999)]. Extrapolation of the excess entropy $S^{ex}$ of liquid water using a power law $C_p$ for the liquid and the experimental heat capacity of ice (available in the CRC handbook of Chemistry and Physics) results in $S^{ex}(T_o)=0$ for a $T_o$ within 1 K of $T_s$, so $T_o$ is around 225 K. This locus for the lower limit of metastability agrees with the temperature of maximum crystallization rate predicted from the thermodynamic and diffusional data for water, 224.5 K (see Figure 2b and supplementary discussion A).

The mean first passage time analysis of the trajectories at 202 K (Supplementary Figure 5b) already evinces a lack of well-defined plateau. This implies that the time scales of growth and nucleation are comparable and –according to Wedekind et al. [*J. Chem. Phys.*, 131, 114506 (2009); *J. Chem. Phys.*, 126, 134103 (2007)]- that the barriers are comparable to the thermal energy $RT$. It is not trivial, however, to extract the actual value of the nucleation barriers from the simulations, because knowledge of the true nucleation coordinate is needed to avoid an overestimation of the barrier. The expression for the nucleation barrier of equation s.2 assumes that the radius of the nucleus is the proper reaction coordinate. An estimation of the barrier at 202 K using equation s.2 assuming a spherical nuclei and with radius as reaction coordinate and that the value of the surface tension is the same as measured around the melting temperature leads to a barrier of ~45 $RT$ at 202 K. The barrier evaluated from equation s.2 becomes 31 $RT$ if we use the scaling of the surface tension with the molar enthalpy of melting proposed by Turnbull [*J. Appl. Phys.* 21, 1022 (1950)]. This is probably still an overestimation, because the time scale of nucleation at 202 K is just 5 ns (assuming the fastest pre exponential factor $\tau_x = h/kT \exp(\Delta G^{\#}/RT)$ and knowing that at 202 K $\tau_x$ = 5 ns predicts $\Delta G^{\#}/RT \approx 11$). We estimate that the experimental barrier for ice nucleation at 230 K is at most twice the nucleation barrier for the mW model at 202 K; this ratio was determined using equation s.2 and the ratio of the experimental $\rho G^{ex}$ extrapolated to 230 K using the equation of Koop et. al. [*Nature*, 406, 611-614 (2000)] (dashed cyan line in Figure 1c) and the $\rho G^{ex}$ in the simulations at 205 K. This implies that the barrier of nucleation is already relatively low just below the experimental $T_H$.

Kauzmann's solution of the entropy paradox, however, did not consider the pre exponential factors for the relaxation and crystallization times. If both pre exponential factors are proportional to the same characteristic time of material transport, then the time of relaxation would not be strictly longer than the time of crystallization, although, as explained by Cavagna [*Physics Reports*, *476*, 51-124 (2009)], even if the relaxation and crystallization times were comparable, it would be insufficient to define a true metastable equilibrium liquid state. Tanaka has argued that the decoupling of the translational diffusion and viscosity in supercooled liquids ensures that on approaching $T_o$ the relaxation time (proportional to the viscosity) becomes much longer than the crystallizaton time (inversely proportional to the translational diffusion, see equation s.4), and thus the supercooled liquid cannot be equilibrated already before reaching $T_o$, avoiding the paradox of an *equilibrated* liquid with entropy lower than the crystal [Tanaka, *Phys. Rev. E.*, *68*, 011505 (2003)]. Below we apply Tanaka's arguments to water.

The lower limit of metastability, the kinetic spinodal, is reached when the time scale of crystallization $\tau_x$ becomes comparable to the time scale of relaxation $\tau_\alpha$ of the liquid. The liquid relaxation is proportional to the viscosity while the crystallization rate is proportional to the time of material transport across the interface of the growing nucleus, $\tau_t$, inversely proportional to the translational diffusion (see [Ngai et al., *J. Chem. Phys.*, *112*, 1887-1892 (2000)] for experimental evidences that crystallization follows the translational diffusion; and [Stevenson and Wolynes, *J. Phys. Chem. A*, 115, 3713 (2011)] for a theoretical foundation of the decoupling of the alpha relaxation and the crystallization rate in deeply supercooled liquids). Tanaka has argued that the solution of the Kauzmann paradox has to account for the fact that the translational diffusion and the viscosity decouple in supercooled liquids [Tanaka, *Phys. Rev. E.*, *68*, 011505 (2003)]:

$$\frac{\tau_x}{\tau_\alpha} = \frac{\tau_t}{\tau_\alpha} g(\phi,T) = \tau_\alpha^{n-1} g(\phi,T)/T \tag{s.6}$$

where *n* is the fractional exponent of the fractional Stokes-Einstein (FSE) relation that measures the decoupling of translational diffusion and structural relaxation (viscosity $\eta$) of the supercooled liquid and the explicit expression for $g(\varphi,T)$ is given in equation s.4. The upper term of $g(\varphi,T)$ in s.4 contains the slowdown of the crystallization due to the nucleation barrier; that barrier decreases with temperature and –according to the simulations- becomes comparable to $RT$ for temperatures just below $T_H$. The denominator of $g(\varphi,T)$ in s.4 measures the accelerating effect of an increase in $G^{ex}$ on the growth rate. We evaluated $(1-\exp(-G^{ex}/RT))^{-3/4}$ using the experimental $G^{ex}$ in Figure 1 and found that it changes little with temperature in the deeply supercooled region, from 2.65 at 240 K to 1.50 at 160 K. This shows that most of the temperature dependence of $g(\phi,T)$ arises from the exponential term containing the nucleation barrier, which becomes very small below the temperature of structural transformation $T_L$. As shown in section A and Figure 2b, the evaluation of the thermodynamic and diffusion contributors to the crystallization time using experimental data for water predicts that the crystallization rate is maximum at 224.5 K.

If the liquid were to follow the standard SE relation (*n* = 1, i.e. *D* proportional to $T/\eta$) the crystallization could never be strictly faster than the relaxation time. Supercooled water, however, follows a fractional SE relation with exponent *n* = 0.62 [Xu et al. *Nature Physics*, *5*, 565-569 (2009)]. This implies that on increasing supercooling the crystallization becomes increasingly faster than the relaxation of the liquid. The diffusion coefficient of water, measured down to 238 K, follows a power law with singular temperature 219.2±2.6 K [Price et al., *J. Phys. Chem. A*, *103*, 448-450 (1999)]. The FSE relation for supercooled water together with the power law behavior for the diffusivity (and the viscosity) implies that $\tau_x/\tau_\alpha$ should be very small already at $T_H$ and that the liquid should reach a limit of metastability at $T_{LLM} > T_s \approx T_o \approx 225$. As the relaxation time has a stronger temperature dependence than the diffusion time, the limit of metastability of real water could be above the 224.5 K predicted to be the nose of the TTT curve; this is the same scenario we report for the simulations with the mW model.

The arguments presented here strongly suggest that real liquid water has a lower limit of metastability between $T_H \approx 232$ K and $T_s \approx 225$ K. The existence of kinetic spinodal in real water close to $T_H$ can only be proved/disproved by performing measurements of crystallization and relaxation times in that region of "no-man's land". State of the art measurements only prove crystallization rates in "no-man's land" at temperatures up to ~155 K, for which molecular diffusivity cannot be measured without interference of crystallization [e.g. Smith and Kay, *Nature*, *398*, 788-791(1999)]. This indicates that 155 K is below the kinetic spinodal of liquid water; methods that allow faster detection of ice crystallization and molecular mobility are needed to determine the lowest limit of metastability of water in experiments.

A kinetic spinodal in bulk water, located between $T_S$ and $T_H$, was anticipated by Kiselev through application of the fluctuation theory of relaxation of metastable states to water employing with the IAPWS-95 equation of state and its extrapolation to the deeply supercooled region [*Int. J. of Thermophys.* **22**, 1421-1433 (2001)]. Kiselev concluded that a liquid-liquid transition consistent with the extrapolated equations of state would be located below the kinetic spinodal for liquid water. Very recently, Limmer and Chandler [arXiv:1107.0337v1 cond-matt (2011)] constructed free energy maps of the mW and ST2 atomistic model of water as a function of the density and the global order parameter $Q_6$ (which can distinguish liquid from ice) and found only two free energy basins: a high density liquid and a low density crystal; but no basin for low-density liquid water. Previous simulations of supercooled water using the ST2 and TIP4P/2005 models have probed regions of the phase diagram below $T_L$ without any evidence of crystallization [Liu et al., *J. Chem. Phys.*, **131**, 104508 (2009); Abascal and Vega, *J. Chem. Phys.*, **133**, 234502 (2010)]. The lack of crystallization may be due to small system size and/or insufficient time to relax the atomistic systems. It is known that finite size affects the barriers of nucleation in simulations [Wedekind et al., *J. Chem. Phys.*, **125**, 214505 (2006)] and that the nucleation time is maximum for systems of size comparable to the critical nucleus [Cavagna et al., *Phys. Rev. Lett.*, **95**, 115702 (2005)]. The simulation cells in these atomistic water studies contain about 500 molecules, comparable to the number of molecules in the critical ice nuclei. A recent simulation study of the freezing and melting of water in partially filled using the mW model ["Melting and crystallization of ice in partially filled nanopores", Gonzalez Solveyra, de la Llave, Scherlis and Molinero, *J. Phys. Chem. B,* DOI: 10.1021/jp205008w (2011)] shows that when water in the partially filled pore forms a liquid plug with a few hundred molecules, nucleation of ice at 180 K is about 1000 times slower than in the bulk simulations with several thousands water molecules; this suggests that extreme confinement or the use of small cells in simulations may allow for the sampling of liquid water for finite times at temperatures below the liquid transformation temperature. The second important issue in the simulations is the equilibration of the supercooled liquid. The ST2 study was performed with very long Monte Carlo simulations; it is not clear how the MC moves affect the crystallization of the system. The TIP4P/2005 study was performed with MD simulations; the authors verified that the water molecules diffuse at least their molecular diameter in 100 ns; due to the decoupling of structural relaxation and translational mobility, however, if the diffusion time were $\tau_D = 100$ ns and the fractional Stokes-Einstein exponent is $n = 0.62$, then a minimum relaxation time $\tau_D^{1/n} \approx 1.7$ μs maybe needed to attain full structural relaxation of the liquid. More work is needed to establish the size dependence of the limit of metastability. As nucleation becomes increasingly easier for large volumes, it should be expected that an increase in size of the system result in a higher limit of metastability of liquid water. The increase in the LLM, however, would not be more than a few degrees, because the volume term appears in the pre exponent that is overruled by the exponential changes arising from an increase of the free energy barrier for nucleation on decreasing the degree of supercooling [Cavagna, *Physics Reports*, **476**, 51-124 (2009)].

**SUPPLEMENTARY FIGURES AND LEGENDS**

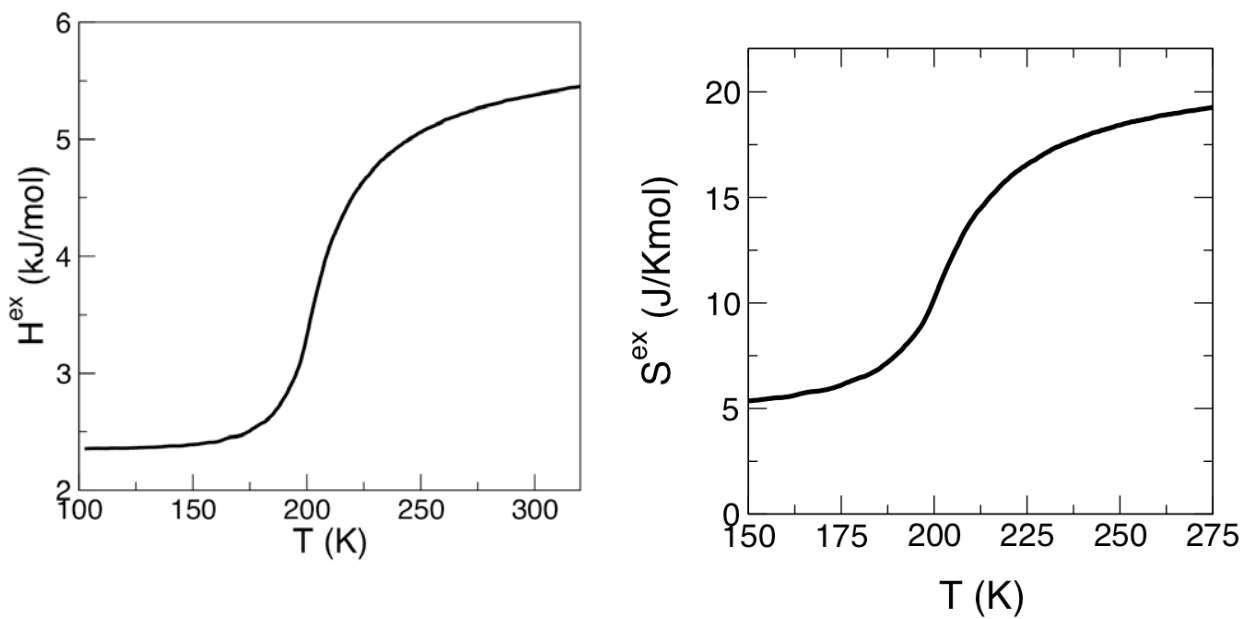

**Supplementary Figure 1. Excess enthalpy $H^{ex}$ and entropy $S^{ex}$ of liquid water with respect to ice.** $H^{ex}$ is the difference between the molar enthalpy of the liquid and the molar enthalpy of ice, both computed on cooling simulations with the mW model at the critical rate of vitrification for the model, 10 K/ns. The values below 202 K do not correspond to an equilibrium liquid or amorphous phase. The excess entropy of LDA reported from the analysis of vapor pressure in experiments at 150 K is even lower than in the simulations, 1.6±1 J/Kmol [Smith et al. *J. Phys. Chem. A.* 115, 5908 (2011)] and suggests that the crystallinity of LDA in experiments could be even larger than the 5% detected by x-ray and calorimetry analysis.

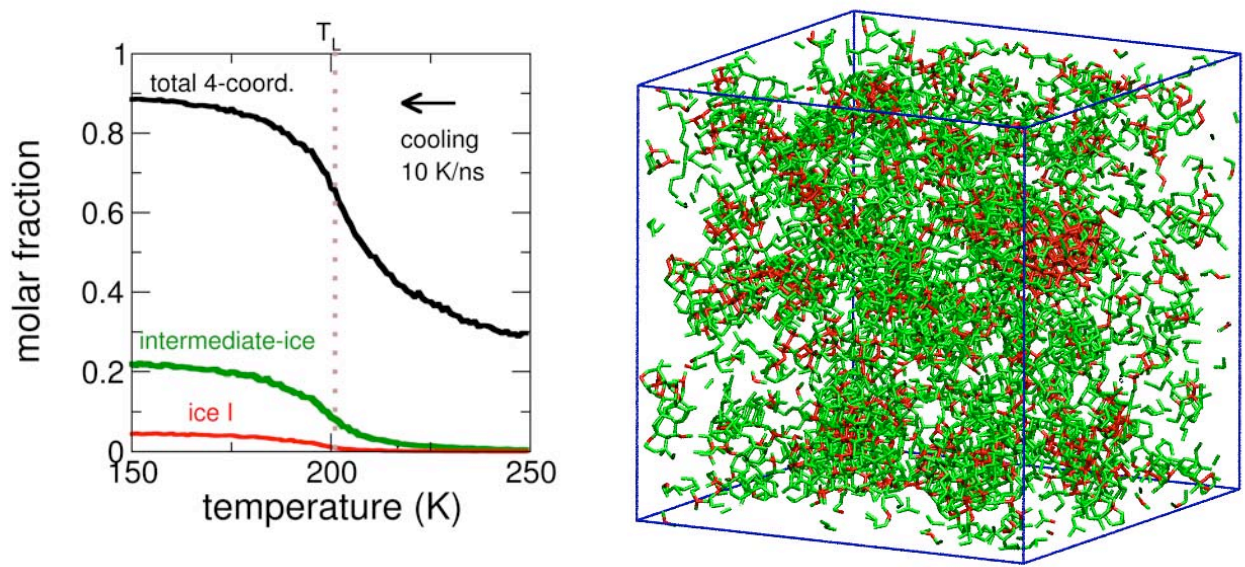

**Supplementary Figure 2. Development of ice and intermediate-ice during the hyperquenching of liquid water.** Left: Fraction of ice I (red), intermediate-ice (green) and total number of 4-coordinated water molecules (black) as a function of temperature on a cooling simulation at 10 K/ns. Right: the resulting structure, LDA, contains small ice crystallites (red) and dense threads of intermediate-ice. The threads of intermediate-ice may correspond to the "gossamer percolative network" of nanocrystallites predicted to form by the random first order transition theory (RFOT) of Stevenson and Wolynes [J. Phys. Chem. A., 115, 3713-3719 (2011)] at temperatures for which the length scale for motions relevant to the α-relaxation is larger than the critical nucleus size for crystallization.

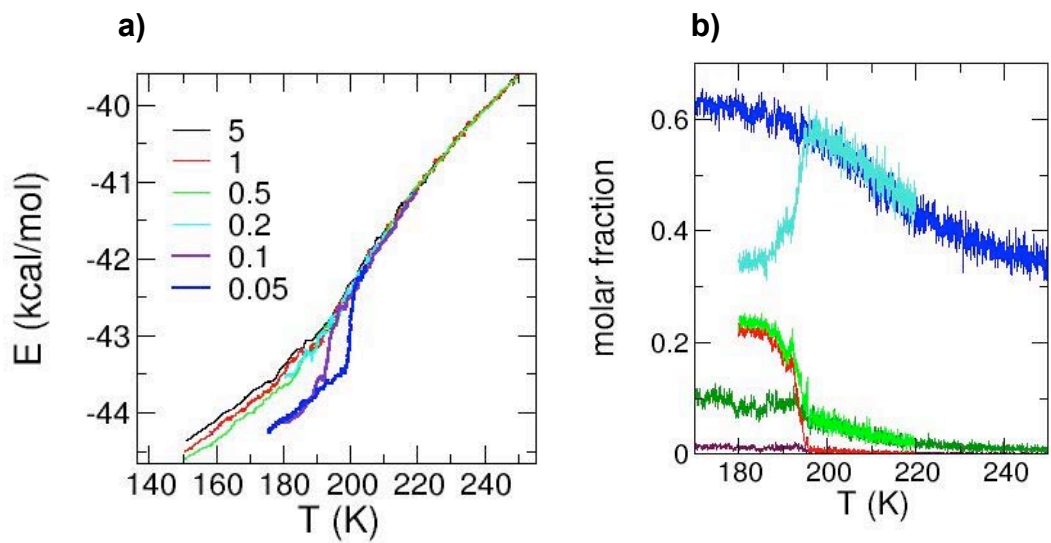

**Supplementary Figure 3. Energy and structural transformation on hyperquenching of water confined in a nanopore. (a)** Enthalpy of water confined in the 3 nm diameter cylindrical nanopore as it is cooled at the rates indicated in the inset, in units of K/ns. On decreasing the cooling rate, water first crystallizes at 196 K [in agreement with the results of Moore et al., *Phys. Chem. Chem. Phys.*, *12*, 4124-4134 (2010)], just 25 K below the melting temperature for water in the pore, $T_m = 220\pm 3$ K [Moore et al. op. cit. above and Gonzalez Solveyra, de la Llave, Scherlis and Molinero, J. *Phys. Chem. B*, DOI: 10.1021/jp205008w (2011)]. The results show an approach of freezing and melting lines as the radius of the nanopore decreases, an observation well established in experiments (see for example [Jaehnert et al., *Phys. Chem. Chem. Phys.*, *10*, 6039-6051 (2008)]). **(b)** Evolution of the structure of water in the pore as it is cooled at 1 and 0.1 K/ns. The lines show the molar fractions of ice (maroon at 1 K/ns and red at 0.1 K/ns), intermediate-ice (dark green at 1 K/ns and light green at 0.1 K/ns) and four-coordinated liquid (blue at 1 K/ns and cyan at 0.1 K/ns). The fraction of four-coordinated liquid increases already at temperatures above the melting point, and has a maximum rate of change around 205 K. As observed in bulk water, the fraction of intermediate-ice increases below 220K, pre announcing the limit of stability of liquid water that can only be studied out of equilibrium below $T_L$. The flat maroon line shows that no ice crystals form at 1 K/ns. At 0.1 K/ns the formation of ice (red line) is accompanied by a depletion of four-coordinated liquid, from which it forms.

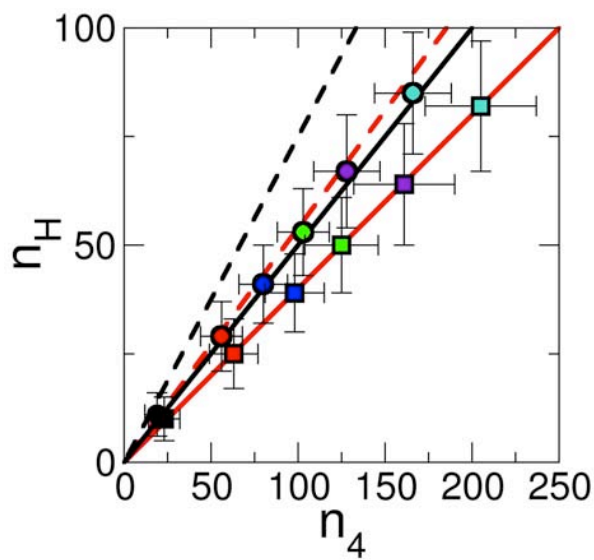

**Supplementary Figure 4. Local environment of ice nuclei is enriched in four-coordinated liquid.** Amount of four-coordinated, $n_4$ and higher-coordinated, $n_H$ molecules in contact with the ice nuclei at 205 K (circles) and 200 K (squares). Data is separated by nuclei sizes, for nuclei containing 10-50 molecules (black), 51-100 (red), 101-150 (blue), 151-200 (green), 201-300 (purple) and 301-400 molecules (turquoise). The average nuclei wetting layer at 205 K (solid black line) has 50% higher fraction of 4-coordinated liquid water molecules than the average measured at that temperature (dashed black line) over all the molecules in the system. At 200 K, the enhancement of 4-coordinated liquid in the wetting layer of the nuclei is lower, 35%, but the fraction of 4-coordinated liquid in the wetting layer is larger than at 205 K. These results suggest that the fraction of four-coordinated molecules around the ice nuclei may be a relevant reaction coordinate for the crystallization of ice.

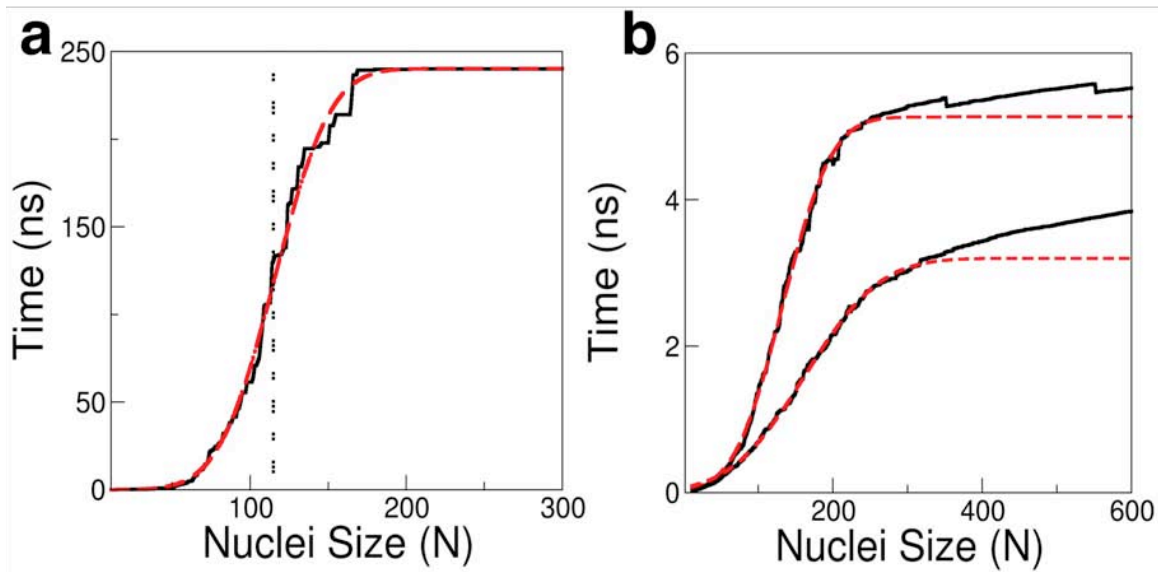

**Supplementary Figure 5. Mean First Passage Time Analysis of Ice Nucleation**. (a) Mean first passage times for the nucleation of ice in the simulations at 208 K (full black line) and the fit to equation 2 of methods (red dashed line). Vertical dashed line is the inflection point of the curve, which corresponds to the critical nucleus size. Uncertainty in the critical nucleus size is determined from the standard deviation of the mean first passage times. The plateau at large (greater than 200) ice nucleus size indicates the nucleation time. (b) Mean first passage time for ice nucleation in the simulations at 202 K (upper curves) and 200 K (lower curves); same notation as in (a). A plateau is not found at large cluster sizes for temperatures below $T_{LL}$ = 202 K, indicating that the growth and nucleation are concurrent and the barriers for nucleation are comparable to the thermal energy $RT$.

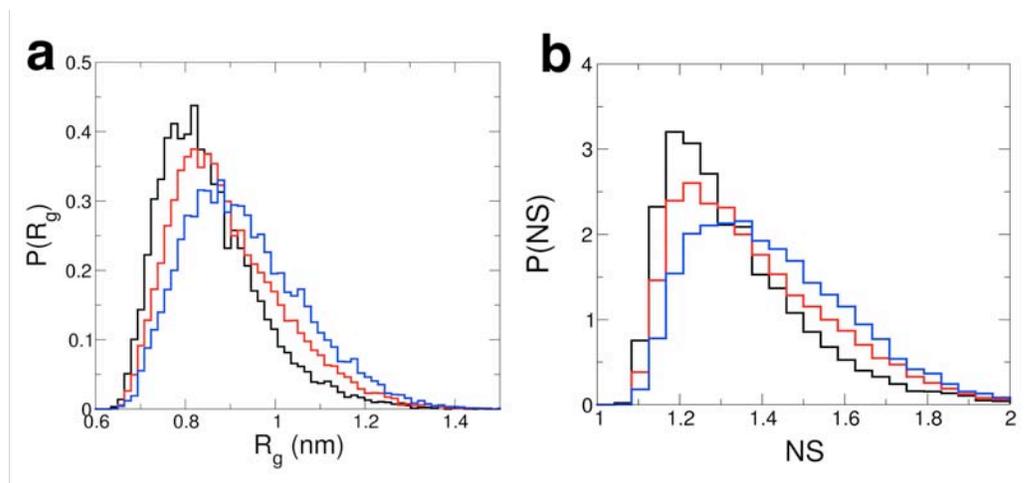

**Supplementary Figure 6. Distribution of radius of gyration $R_g$ and nonsphericity NS for ice nuclei of size around the critical size N\* deduced from mean first passage time analysis.** NS is defined as the ratio between the Rg of the nucleus and the one of an ice sphere with the same number of water molecules. Panel (a) shows a histogram of the $R_g$ values for all sized nuclei with 70-130 water molecules at three different temperatures, 205 K (black), 202 K (red) and 200 K (blue). For these same temperatures, panel (b) shows the histogram of the NS values for all these nuclei.

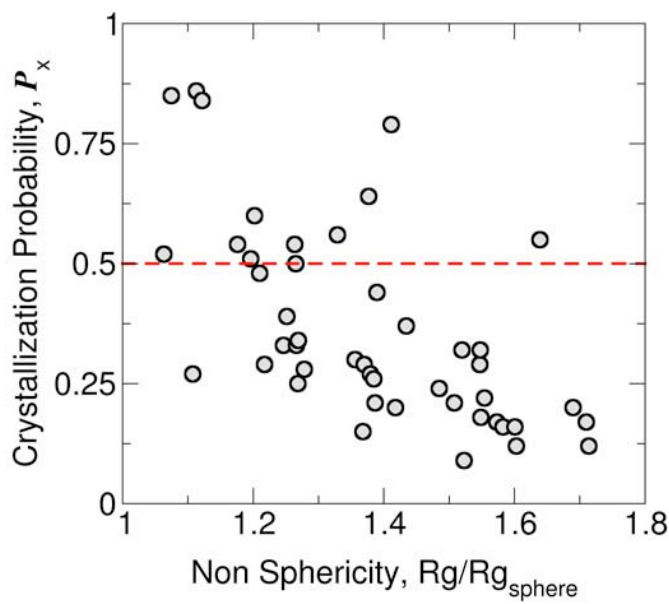

**Supplementary Figure 7. Crystallization probability of ice nuclei of varied non sphericity.** Crystallization probability (aka commitment probability) of individual nuclei that are of critical size, N* 90 to 100 molecules at 205 K, according to the MFPT analysis using $N$ as the reaction coordinate for the crystallization. Only nuclei with $P_x$ close to 0.5 are critical. The commitment probability shows a broad spread of values suggesting that other reaction coordinates, such as the structure of the liquid wetting the nuclei, are relevant to define the transition states of ice crystallization.